\documentclass[preprints,article,accept,moreauthors,pdftex]{Definitions/mdpi} 

\usepackage{subcaption}
\usepackage{comment}
\usepackage{tikz}
\usepackage{tabularx}
\usepackage{xcolor}
\usetikzlibrary{shadows,arrows}

\pgfdeclarelayer{background}
\pgfdeclarelayer{foreground}
\pgfsetlayers{background,main,foreground}
 
\tikzstyle{materia}=[draw, fill=blue!20, text width=6.0em, text centered, minimum height=1.5em,drop shadow]
\tikzstyle{Block} = [materia, text width=6em, minimum width=10em,
  minimum height=3em, rounded corners, drop shadow, text centered]
\tikzstyle{DimBlock} = [materia, text width=10em, minimum width=10em,
  minimum height=3em, rounded corners, drop shadow, text centered]
\tikzstyle{texto} = [above, text width=6em, text centered]
\tikzstyle{line} = [draw, thick, color=black!80, -latex']

\usepackage{verbatim}
\usepackage{forest}
\usepackage{pgfplots}
\pgfplotsset{compat=1.16}

\pgfplotsset{every axis/.style={scale only axis}}

\firstpage{1} 
\pubvolume{xx}
\issuenum{1}
\articlenumber{5}
\pubyear{2021}
\copyrightyear{2021}
\history{Received: date; Accepted: date; Published: date}

\Title{Introduction and analysis of a method for the investigation of QCD-like tree data}

\Author{Marko Jercic $^{1}$*, Ivan Jercic $^{1}$ and Nikola Poljak$^{1}$}

\AuthorNames{Marko Jercic, Ivan Jercic, Nikola Poljak}

\address[1]{%
$^{1}$ \quad Department of Physics, Faculty of Science, University of Zagreb}

\corres{Correspondence: mjercic@phy.hr}

\abstract{The properties of decays that take place during jet formation cannot be easily deduced  from  the  final  distribution  of particles in a detector. In this work, we first simulate a system of particles with well defined masses, decay channels, and decay probabilities. This presents the ,,true system'' for which we want to reproduce the decay probability distributions. Assuming we only have the data that this system produces in the detector, we decided to employ an iterative method which uses a neural network as a classifier between events produced in the detector by the ,,true system'' and some arbitrary ,,test system''. In the end, we compare the distributions obtained with the iterative method to the ,,true'' distributions.}


\begin{document}

\section{Introduction}
The properties of particle interactions determine the evolution of a quantum chromodynamical (QCD) system. Thorough understanding of these properties can help answer many fundamental questions in physics, such as the origin of the Universe or the unification of forces. This is one of the important reasons to collect data with particle accelerators, such as the Large Hadron Collider (LHC) at CERN. However, when collecting this data, we only register complex signals of high dimensionality which we can later interpret as signatures of final particles in the detectors. This interpretation stems from the fact that we, more or less, understand the underlying processes that produce the final particles.

In essence, from the all the particles produced in a collision at the accelerator, only the electron, the proton, the photon and the neutrinos are stable and can be reconstructed with certainty, given that you have the proper detector. Other particles are sometimes also directly detected, given that they reach the active volume of the detector without first decaying. These include muons, neutrons, charged pions and charged kaons. On the other hand, short-lived particles will almost surely decay before reaching the detector and we can only register the particles they decay into.

A similar situation arises with quarks, antiquarks and gluons, the building blocks of colliding nuclei. When a high energy collision happens, a quark within a nucleus behaves almost as if it doesn't interact with neighbouring particles, because of a property called asymptotic freedom. If it is struck with a particle from the other nucleus, it can be given sufficient momentum pointing outwards from the parent nucleus. However, we know that there are no free quarks in nature and that this quark needs to undergo a process called hadronisation. This is a process in which quark-antiquark pairs are generated such that they form hadrons. Most of the hadrons are short-lived and they decay into other, more stable, hadrons. The end result of this process is a jet of particles whose average momentum points in the direction of the original outgoing quark. Unfortunately, we don't know the exact quark, nor gluon, decay properties, which serves as a motivation for this work.

The determination of these properties is a long standing problem in particle physics. To determine them, we turn to already produced data and try to fit decay models onto them. With every new set of data our understanding changes. This is evident from the fact that we want to simulate a collision event, we can obtain, on average, slightly different results with different versions of the same tool \cite{pythia}. Therefore, even though simulation tools are regularly reinforced with new observations from data, we can not expect the complete physical truth from them.

Instead of trying to perform direct fits to data, we propose the use of machine learning methods to determine the decay properties. In fact, the onset of these methods is already hinted in the traditional approach, since a multivariate fit of decay models to data is already a form of a machine learning technique. It is only natural to extend the existing methods since we can't rely entirely on simulated data. In this work, we develop an interpretable model by first simulating a system of particles with well defined masses, decay channels, and decay probabilities. We take this to be the ,,true system'', whose decay properties we pretend not to know and want to reproduce. Mimicking the real world, we assume to only have the data that this system produces in the detector. Next, we employ an iterative method which uses a neural network as a classifier between events produced in the detector by the ,,true system'' and some arbitrary ,,test system''. In the end, we compare the distributions obtained with the iterative method to the ,,true'' distributions.

This paper is organized as follows: in the Materials and methods section we describe the developed artificial physical system and the algorithm used to recover underlying probability distributions of the system. Also, we present in detail the methodology used to obtain the presented results. In the Results section we present our findings to see whether our hypothesis holds true. We conclude the paper with the Discussion section...

\section{Materials and methods}

The code used for the development the particle generator, the neural network models and the calculations is written in the the Python programming language using the Keras module with the TensorFlow2 backend \cite{keras}. The calculations were performed using a standardized PC setup equipped with an NVIDIA Quadro p6000 graphics processing unit.

\subsection{The physical system}
In particle physics, jets are detected as collimated streams of particles. The jet production mechanism is in essence clear: partons from the initial hard process undergo the fragmentation and hadronization processes. In this work, we develop a simplified physical model in which the fragmentation process is modeled as cascaded $1 \rightarrow 2$ independent decays of partons with a constant number of decays. This way, any single jet can be represented as a perfect binary tree of depth $N$, corresponding to $2^N$ particles in the final state. Since the initial parton properties are set, jets can be described by $2^N - 1$ decay parameters. We represent each decay of a mother parton of mass $M$ by four real numbers $(\frac{m_1}{M}, \frac{m_2}{M}, \theta, \phi)$, where $m_1$ and $m_2$ are the masses of the daughter particles and $\theta$ and $\phi$ are the polar and azimuthal angle of the lighter particle, as measured from the rest frame of the mother particle. For simplicity we make all the decays isotropic, which isn't necessarily true in real processes. To fully define our physical system we set a decay probability distribution function $p(m_1, m_2 | M)$, the details of which are given in the following subsection. The aim of our proposed algorithm is to recover these underlying probability distributions, assuming we have no information on them, using only a dataset consisting of jets described with final particles' four-momenta, as one would get from a detector.   

\subsection{Particle generator}
\label{ParticleGenerator}
To generate the jets, we developed an algorithm where we take a particle of known mass that undergoes three successive decays. We consider only the possibility of discrete decays, in the sense that the decay product masses and decay probabilities are well defined. We consider a total of 10 types of particles, labelled A -- J, which can only decay into each other. The masses and the decay probabilities of these particles are given in Table \ref{TableParticles}. In this scenario, the ,,decay probabilities'' $p$ are given by the ratios of decay amplitudes. Thus, the total sum of the probabilities for a given particle to decay into others has to be one, and the probabilities describe the number of produced daughters per $N$ decays, scaled by $1/N$.

\vskip 5mm
\begin{table}[h!t!]
    \centering
\begin{tabular}{|c|c|c|c|c|c|c|c|c|c|c|c|}
\hline
\multicolumn{2}{|c|}{particle} & \multicolumn{2}{|c|}{A} & \multicolumn{2}{|c|}{B} & \multicolumn{2}{|c|}{C} & \multicolumn{2}{|c|}{D}& \multicolumn{2}{|c|}{E} \\\hline
\multicolumn{2}{|c|}{mass} & \multicolumn{2}{|c|}{0.1} & \multicolumn{2}{|c|}{0.6} & \multicolumn{2}{|c|}{1.3} & \multicolumn{2}{|c|}{1.9}& \multicolumn{2}{|c|}{4.4} \\\hline
\multicolumn{2}{|c|}{$p$ / channel} & 1 & A & 0.7 & B & 1 & C & 0.3 & A+C & 0.6 & C+C \\\hline
\multicolumn{2}{|c|}{} &  &  & 0.3 & A+A &  &  & 0.3 & A+A & 0.4 & E \\\hline
\multicolumn{2}{|c|}{} &  &  &  &  &  &  & 0.4 & D & &  \\\hline\hline
\multicolumn{2}{|c|}{particle} & \multicolumn{2}{|c|}{F} & \multicolumn{2}{|c|}{G} & \multicolumn{2}{|c|}{H} & \multicolumn{2}{|c|}{I}& \multicolumn{2}{|c|}{J} \\\hline
\multicolumn{2}{|c|}{mass} & \multicolumn{2}{|c|}{6.1} & \multicolumn{2}{|c|}{8.4} & \multicolumn{2}{|c|}{14.2} & \multicolumn{2}{|c|}{18.1}& \multicolumn{2}{|c|}{25} \\\hline
\multicolumn{2}{|c|}{$p$ / channel} & 0.5 & A+A & 0.9 & B+B & 0.6 & D+D & 1 & F+G & 0.5 & F+I \\\hline
\multicolumn{2}{|c|}{} & 0.5 & B+C & 0.1 & A+F & 0.25 & D+E &  &  & 0.4 & G+H \\\hline
\multicolumn{2}{|c|}{} &  &  &  &  & 0.15 & E+F &  &  & 0.1 & E+E \\\hline
\end{tabular}
\caption{Allowed particle decays in the discrete model. The designation $p$/channel shows the probability that a mother particle will decay into specific daughters.}
\label{TableParticles}
\end{table}
\vskip 5mm

Particles A--E are set to be long lived and can thus be detected in the detector, which only sees the decay products after several decays. This can be seen in table \ref{TableParticles} as a probability for a particle to decay into itself. In this way, we assure two things: first, that we have stable particles and second, that each decay in the binary tree is recorded, even if it is represented by a particle decaying into itself. Particles A and C are completely stable, since they only have one ,,decay'' channel, in which they decay back into themselves. On the other hand, particles F--I are hidden resonances: if one of them appears in the i-th step of the decay chain, it will surely decay into other particles in the next, (i+1)-th step of the chain.

\begin{figure}[h!t!]
  \centering   
\begin{forest}
for tree={
      grow=east,
      edge={->},
      parent anchor=east,
      child anchor=west,
      s sep=1pt,
      l sep=1cm
     },
[J
 [F
  [A[A]]
  [A[A]]
 ]
 [I
  [F
   [B]
   [C]
  ]
  [G
   [B]
   [B]
  ]
 ]
]
\end{forest}
\caption{An example of the operation of the discrete jet generator. The mother particle J decays into particles I and F. According to decay probabilities, this happens in half the generated jets. The daughter particles subsequently decay two more times, leaving only stable, detectable particles in the final state.}
\label{FigRaspadi}
\end{figure}
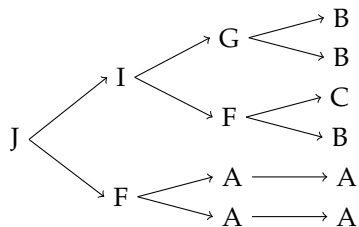

To create a jet, we start with particle J, which we call the mother particle, and allow it to decay in one of the decay channels. Each of the daughter particles then decays according to their decay channels, and this procedure repeats a total of 3 times. In the end, we obtain a maximum of 8 particles from the set A--E, with known momenta as measured from the rest frame of the mother particle. An example of a generated jet is given in Fig.\ref{FigRaspadi}.

\subsection{Introduction to the algorithm}
Let's assume we have two distinct datasets: one that consists of samples from a random variable X distributed with an unknown probability density $p(x)$, which we call the ,,real'' dataset, and the other, which consists of samples from a random variable Y distributed with a known probability density $q(x)$, which we call the ,,test'' dataset. We would like to perform a hypothesis test between $H_{0}:p = p(x)$ and $H_{1}:p = q(x)$ using a likelihood-ratio test. The approach we use follows earlier work employs the Neyman–Pearson lemma \cite{NNNP1, NNNP2, NNNP3}. This lemma states that the likelihood ratio, $\Lambda$, given by:

\begin{equation}
    \Lambda (p \mid q)\equiv \frac {{\mathcal {L}}(x \mid real)}{{\mathcal {L}}(x \mid test)} = \frac{p(x)}{q(x)}
    \label{NP}
\end{equation}
is the most powerful test at the given significance level \cite{NeyPear}. 

We can obtain an approximate likelihood ratio $\Lambda$ by transforming the output of a classifier used to discriminate between the two datasets. Assume that the classifier is a neural network optimized by minimizing the \textit{crossentropy} loss. In this case, the network output gives the probability of $x$ being a part of the real dataset $C_{nn}(x) = p(real \mid x)$ \cite{NNProbability}. If the datasets consist of the same number of samples, we can employ the Bayes' theorem in a simple manner:

\begin{eqnarray}
p(real \mid x) &=& \frac{p(x \mid real)p(real)}{p(x \mid real) p(real)+p(x \mid test)p(test)} \nonumber \\
&=& \frac{p(x \mid p_{\textrm{real}})}{p(x \mid real)+p(x \mid test)} = \frac{\Lambda}{\Lambda+1}\,.
\label{Bayes}
\end{eqnarray}
A simple inversion of Eq.\ref{Bayes} gives:

\begin{equation}
\Lambda = \frac{p(x)}{q(x)}  = \frac{C_{\textrm{NN}}(x)}{1 - C_{\textrm{NN}}(x)},
\end{equation}

\begin{equation}
p(x)  = \frac{C_{\textrm{NN}}(x)}{1 - C_{\textrm{NN}}(x)} q(x).
\label{pq}
\end{equation}
Therefore, in ideal conditions, the unknown probability density $p(x)$ describing the real dataset can be recovered with the help of the known probability density $q(x)$ and a classifier, using \ref{pq}. It must be noted that \ref{pq} is strictly correct only for optimal classifiers, which are unattainable. In our case, the classifier is optimized by minimizing the \textit{crossentropy} loss defined by:

\begin{equation}
L = -\frac{1}{n}\sum_{i=1}^{n}\left[y(x_i)\ln C_{\textrm{NN}}(x_i) + (1-y(x_i))\ln (1-C_{\textrm{NN}}(x_i)) \right]\,,
\end{equation}
where $y(x_i)$ is 1 if $x_i$ is a part of the real dataset, and 0 if $x_i$ is a part of the test dataset. We can safely assume that the final value of loss of the suboptimal classifier is greater than the final value of loss of the optimal classifier:
\begin{equation}
L_{\textrm{optimal}} < L < \ln{2} \,.   
\end{equation}
The value of $\ln 2$ is obtained under the assumption of the \textit{worst} possible classifier. To prove our findings, in the next step we regroup the sums in the loss function in two parts, corresponding to the real and the test distributions:

\begin{equation}
     -\frac{1}{n}\sum_{i \in  real}\ln C_{\textrm{NN}}^{\textrm{optimal}}(x_i) <  -\frac{1}{n}\sum_{i \in  real}\ln C_{\textrm{NN}}(x_i) < -\frac{1}{n}\sum_{i \in  real}\ln \frac{1}{2},
     \label{Lreal}
\end{equation}

\begin{equation}
     -\frac{1}{n}\sum_{i \in  test}\ln\left[1 - C_{\textrm{NN}}^{\textrm{optimal}}(x_i) \right]<  -\frac{1}{n}\sum_{i \in  test}\ln\left[1 - C_{\textrm{NN}}(x_i)\right]  < -\frac{1}{n}\sum_{i \in  test}\ln \frac{1}{2}.
     \label{Ltest}
\end{equation}
After expanding inequality \ref{Lreal} we obtain:

\begin{equation}
     -\frac{1}{n}\sum_{i \in  real}\ln \left[ \frac{C_{\textrm{NN}}^{\textrm{optimal}}(x_i)}{1 - C_{\textrm{NN}}^{\textrm{optimal}}(x_i)}\right] <  -\frac{1}{n}\sum_{i \in  real}\ln \left[\frac{C_{\textrm{NN}}(x_i)}{1 - C_{\textrm{NN}}(x_i)}\right] < -\frac{1}{n}\sum_{i \in  real}\ln 1.
     \label{Expanded}
\end{equation}
According to Eq.\ref{pq}, we can recover the real probability density $p(x)$ when using the optimal classifier. However, if one uses a suboptimal classifier, a slightly different probability density $p'(x)$ will be calculated. Since the ratios that appear as arguments of the logarithms in Eq.\ref{Expanded} correspond to distribution ratios, it follows that:

\begin{equation}
     -\frac{1}{n}\sum_{i \in  real}\ln \left[ \frac{p(x_i)}{q(x_i)}\right] < -\frac{1}{n}\sum_{i \in  real}\ln \left[ \frac{p'
     (x_i)}{q(x_i)}\right] < -\frac{1}{n}\sum_{i \in  real}\ln 1.
\end{equation}
After some simplification this becomes:
\begin{equation}
     \sum_{i \in real} \ln p(x_i) > \sum_{i \in real} \ln p'(x_i) > \sum_{i \in real} \ln q(x_i).
\label{proof1}
\end{equation}
If an analogous analysis is carried out for inequality \ref{Ltest} we get:
\begin{equation}
     \sum_{i \in test} \ln p(x_i) < \sum_{i \in test} \ln p'(x_i) < \sum_{i \in test} \ln q(x_i).
\label{proof2}
\end{equation}
From this, it can be seen that probability density $p'(x)$ is on average closer to the real probability density $p(x)$ than to the test probability density $q(x)$. In a realistic case,  Eq.\ref{pq} can't be used to completely recover the real probability density $p(x)$. However, it can be used in an iterative method; starting with a known distribution $q(x)$, we can approach the real distribution more and more with each iteration step.

\subsection{A simple example}
\begin{figure}[h!t!]
    \centering
        \begin{subfigure}{0.49\textwidth}
        \includegraphics[width=\linewidth]{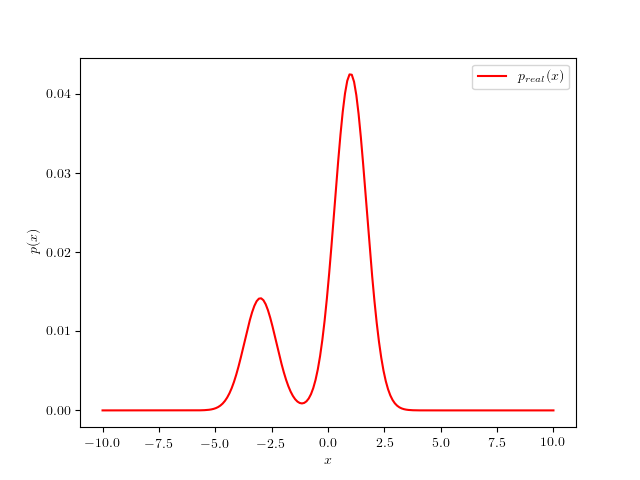}
        \end{subfigure}
        \begin{subfigure}{0.49\textwidth}
        \includegraphics[width=\linewidth]{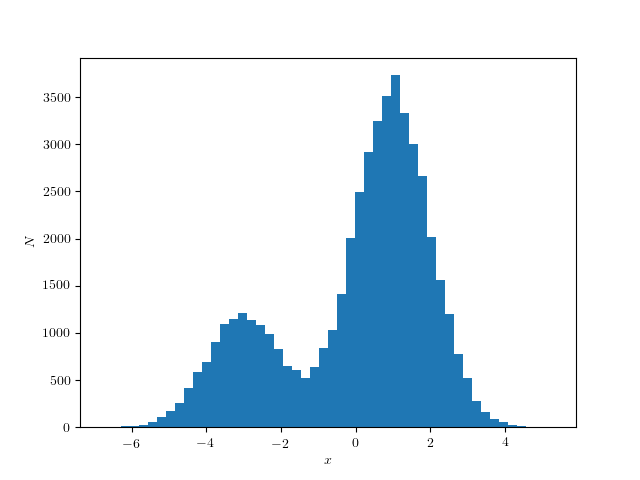}
        \end{subfigure}
\caption{(\textbf{a}) The normalized probability density for the example, given by Eq. \ref{eqpreal}. (\textbf{b}) A histogram of values sampled from the set generated by the same equation.}
\label{hsimple}
\end{figure}

Let us illustrate the recovery of an unknown probability density by using a classifier on a simple example. We start with a set of 50 000 real numbers generated from a random variable with a probability density given by
\begin{equation}
 p_{\textrm{real}}(x) = \frac{1}{4} \mathcal{N}(-1,1) + \frac{3}{4}\mathcal{N}(3,1)\,,
 \label{eqpreal}
 \end{equation}
where $\mathcal{N}(\mu,\sigma^2)$ denotes a normal distribution. A histogram of values in this set is shown in \ref{hsimple}. Let's now assume we don't know $p_{\textrm{real}}(x)$ and want to recover it using the procedure outlined in the previous subsection. This set will be denoted as the ,,real'' dateset and the underlying probability density will be denoted as the ,,real'' probability density.    

\begin{figure}[h!t!]
\centering
\includegraphics[width=15cm]{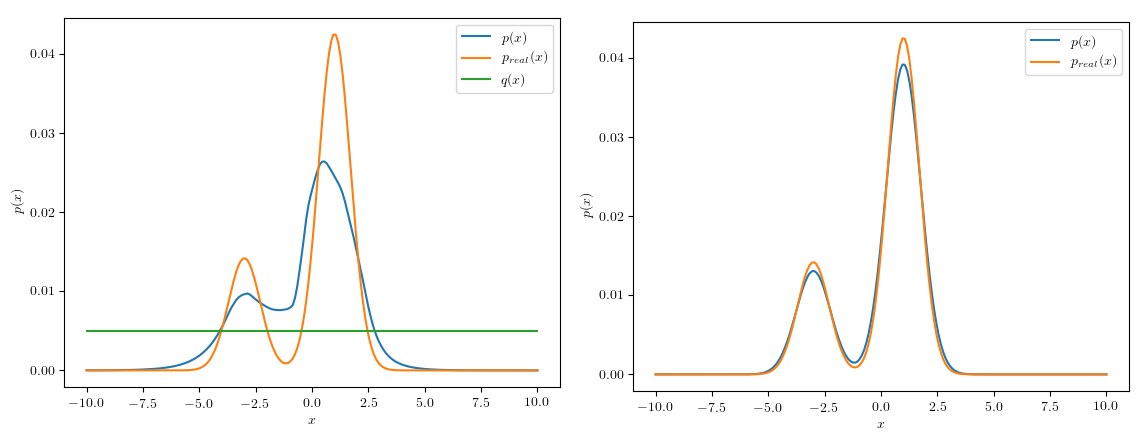}
\caption{The calculated $p_{\textrm{calculated}}(x)$ (blue line) compared to the real probability density $p_{\textrm{real}}$(x) (orange line). (\textbf{a}) The left panel shows the comparison after one iteration of the algorithm, alongside the starting ,,test'' distribution (green line).(\textbf{b}) The right panel shows the comparison after the 5th iteration.}
\label{nn_simple_0}
\end{figure}

To construct the ,,test'' dataset, we generate values with a uniform probability density in the interval $\left[-10,10 \right]$. Finally, we construct a simple neural network which is used as a classifier that distinguishes the examples from the real dataset from examples from the test dataset. The classifier we use is a simple \textit{feed-forward} neural network with 100 hidden units using a ReLU activation function. The activation function of the final neural network output is the \textit{sigmoid} function, which we use to constrain the output values to the interval [0,1]. After the classifier is trained to discriminate between the two datasets by minimizing the \textit{binary crossentropy} loss, we evaluate its output at 200 equidistant points between -10 and 10. Using Eq.\ref{pq}, the probability distribution $p_{\textrm{calculated}}(x)$ is calculated using the classifier outputs. The calculated $p_{\textrm{calculated}}(x)$ is compared to the real probability density $p_{\textrm{real}}(x)$ and is shown in \ref{nn_simple_0}. 

Although the resulting probability density differs from the real probability density due to the non-ideal classifier, we can conclude that the calculated $p_{\textrm{calculated}}(x)$ is considerably closer to $p_{\textrm{real}}(x)$ than to uniform probability density $q(x)$ used to generate the test dataset. Now, if we use the calculated $p_{\textrm{calculated}}(x)$ to construct a new test dataset and repeat the same steps, we can improve the results even more. This procedure can therefore iteratively improve the resemblance of $p_{\textrm{calculated}}(x)$ to $p_{\textrm{real}}(x)$ to the point where the datasets are so similar that the classifier cannot distinguish between them. In this simple example convergence is reached after the 5th iteration, since no significant improvement is observed afterwards. The calculated probability density $p_{\textrm{calculated}}(x)$ after the final iteration is shown in \ref{nn_simple_0} compared to the real distribution $p_{\textrm{real}}(x)$. It is clear that in this case the procedure converges, and we could possibly obtain a better match between $p_{\textrm{calculated}}(x)$ and $p_{\textrm{real}}(x)$ if we used a more optimal classifier.

In essence, a simple histogram could be used in this simple example to determine the underlying probability distribution instead of using the method described above. However, in case of multivariate probability distributions, which can be products of unknown probability distributions, a histogram approach would not prove useful.

\subsection{Extension to jets}
We would now like to apply the described procedure on the datasets that contain jets. Every jet, represented by a binary tree of depth $N$, consists of $2^N-1$ independent decays producing a maximum of $2^N$ particles in the final state. Since all the decays are isotropic in space, a jet can be described with a 4 $\times$ $(2^N-1)$--dimensional vector $\vec{x}$ and a probability distribution function given by: 
\begin{equation}
p\left(\vec{x} \right) = \prod_i^{2^N-1} p(m_1^i, m_2^i | M)p(\theta^i) p(\phi^i),
\label{jet_prob}
\end{equation}
where $i$ denotes the decay index and ($m_1^i$, $m_2^i$, $\theta^i$, $\phi_i$) are the components of the vector $\vec{x}$. Since both angles are uniformly spatially distributed, they contribute to the probability with a simple constant factor. Therefore, when plugging $p\left(\vec{x} \right)$ from Eq.\ref{jet_prob} into Eq.\ref{pq} we can omit angles, since the constant factors will cancel each other out:
\begin{equation}
\prod_i^{2^N-1} p(m_1^i, m_2^i | M) = \frac{C_{NN}(\vec{x})}{1 - C_{NN}(\vec{x})} \prod_i^{2^N-1} q(m_1^i, m_2^i | M).
\label{pq_jets}
\end{equation}
Taking the logarithm of both sides:
\begin{equation}
\sum_i^{2^N-1} \ln p(m_1^i, m_2^i | M) = \ln{C_{NN}(\vec{x})} - \ln({1 - {C_{NN}(\vec{x})}})  + \sum_i^{2^N-1} \ln q(m_1^i, m_2^i | M).
\label{log_pq_jets}
\end{equation}
Unfortunately, we cannot explicitly obtain the probability $p(m_1, m_2 \mid M)$ directly from Eq.\ref{log_pq_jets} without solving a linear system of equations. This task proves to be computationally exceptionally challenging due to the high dimensionality of the dataset. In order to avoid this obstacle, we introduce a neural network $f$ to approximate $\ln p(m_1,m_2|M)$. We can optimize this neural network by minimizing the \textit{mean squared error} applied to the two sides of Eq.\ref{log_pq_jets}. 

\subsection{The 2 Neural Networks (2NN) algorithm}
At this point we are ready to recover the underlying probability distributions from an existing dataset that consists of jets described by the four-momenta of the final particles. We denote the jets from this dataset as ,,real''. The building blocks of the full recovery algorithm are two independent neural networks; the aforementioned classifier $C_{NN}$ and the neural network $f$. Based on the usage of 2 neural networks, we dubbed the algorithm \textit{2NN}. The detailed architectures of both networks are given in Appendix A. 

The workflow of the 2NN algorithm is simple: first we initialize the parameters of both neural networks. Then, we generate a test dataset using the neural network $f$. The test dataset and the real dataset are fed into the classifier network, which produces a set of linear equations in the form of Eq.\ref{log_pq_jets}. We approximate the solution to these by fitting the neural network $f$, which in turn produces a new test dataset. The procedure is continued iteratively until there are no noticeable changes in the difference of the real and test distributions. More detailed descriptions of the individual steps are given in the next subsections.

\subsubsection{Generating the test dataset}

\begin{figure}[h!t!]
\begin{center}
\resizebox{\columnwidth}{!}{%
    \begin{tikzpicture}
    \draw[->,thick,>=stealth] (-1,0) -- (12,0) node[right] {{\huge $a$}};
    \draw[->,thick,>=stealth] (0,-1) -- (0,12) node[left] {{\huge $b$}};
    \draw[dashed,thick] (0,10)--(10,0);
    \draw[dashed,thick] (0,0)--(10,10);
    \node[rotate=45] at (8,8.5) {\huge $a = b$};
    \node[rotate=-45] at (8,2.5) {\huge $a + b = 1$};
    \fill[black!10] (0,0) -- (5,5) -- (0,10) -- cycle;
    \node[] at (2,5) {{\fontsize{40}{60}\selectfont $\Omega$}};
    
    \draw[->,thick,>=stealth] (14,0) -- (27,0) node[right] {{\huge $a$}};
    \draw[->,thick,>=stealth] (15,-1) -- (15,12) node[left] {{\huge $b$}};
    \draw[dashed,thick] (15,10)--(25,0);
    \draw[dashed,thick] (15,0)--(25,10);
    \node[rotate=45] at (23,8.5) {\huge $a = b$};
    \node[rotate=-45] at (23,2.5) {\huge $a + b = 1$};

    \foreach \x in {0,1,2,...,25} {
    \draw[thick] (15+0.2*\x,-0.2+0.2*\x) -- (15+0.2*\x,10.2-0.2*\x);};
    \foreach \y in {0,1,2,...,25} {
    \draw[thick] (15,-0.2+0.2*\y) -- (15+0.2*\y,-0.2+0.2*\y);};
    \draw[thick] (15,-0.2+0.2*26) -- (15+0.2*25,-0.2+0.2*26);
    \foreach \y in {0,1,2,...,25} {
    \draw[thick] (15,5.2+0.2*\y) -- (20-0.2*\y,5.2+0.2*\y);};
\end{tikzpicture}%
}
\caption{(\textbf{a}) The left panel shows the entire allowed probability space of the parameters $a$ and $b$, designated by $\Omega$. Due to conservation laws, $a+b \leq 1$ needs to hold true. To describe our system, we selected the case where $a \leq b$, which we can do without loss of generality. (\textbf{b}) The right panel shows the discretized space $\Omega$, as used to evaluate the partition function.}
\label{prob_space}
\end{center}
\end{figure}
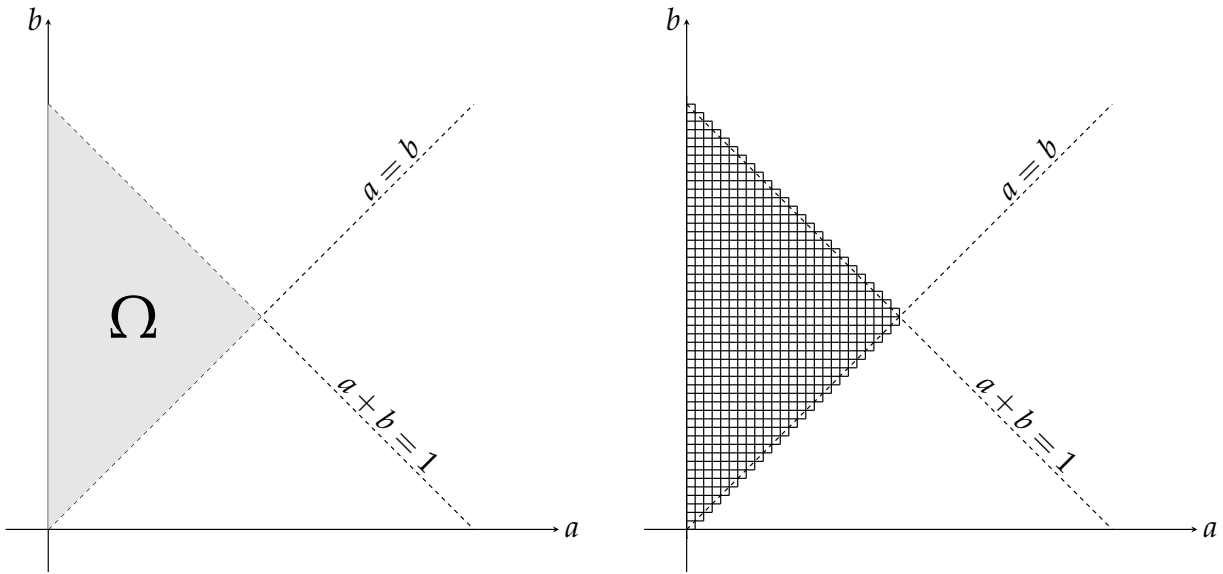

After the parameters of the neural network $f$ are initialized, we need to generate a test dataset of jets with known decay probabilities $q(\vec{x})$. The input of the neural network $f$ is a vector consisting of 3 real numbers: $a = m_1/M$, $b = m_2/M$ and $M$. We denote the output of the neural network with $f(a,b,M)$. Due to conservation laws, the sum $a+b$ needs to be strictly less or equal to 1. We can assume $a \leq b$ without any loss of generality. In order to manipulate with probabilities a partition function:
\begin{equation}
Z(M) = \int_{\Omega} e^{f(a,b,M)} \,\mathrm{d}a \mathrm{d}b
\label{Z}
\end{equation}
needs to be calculated. Here, $\Omega$ denotes the entire probability space and is shown as the gray area in the left panel of Fig. \ref{prob_space}. To calculate the integral in the above expression, the probability space is discretized into 650 equal areas shown in the right panel of Fig.\ref{prob_space}. These areas are obtained by discretizing the parameters $a$ and $b$ into equidistant segments of length 0.2. After the discretization, the partition function $Z(M)$ then becomes:
\begin{equation}
Z(M) \approx \sum_j \sum_{k} e^{f(a_j,b_k,M)} \,.
\label{Z_discrete}
\end{equation}

To generate the jets which form the test dataset, we must generate each decay in the cascading evolution using the neural network $f$. Each of the decays is generated by picking a particular pair of parameters $(a,b)$ from the 650 possible pairs which form the probability space for a given mass $M$. The decay probability is then given by:
\begin{equation}
q(m_1, m_2 \mid M) = \frac{e^{f(a,b,M)}}{Z(M)}\,. 
\label{q}
\end{equation}
After applying this procedure we have a test dataset in which each jet is represented as a list of $2^N$ particles and their four-momenta. For each decay, we also store the pairs $(a^i,b^i)$ as well the corresponding decay probabilities. 

\subsubsection{Optimizing the classifier}
The classifier used in this work is a convolutional neural network. The input to these type network are sets of images. For this purposes, all the jets are preprocessed by transforming the list of particles' four-momenta into jet images. Two 32$\times$32 images are produces for a single jet. In both images the axes correspond to the decay angles $\theta$ and $\phi$, while the pixel values are either the energy of the momentum of the particle found in that particular pixel. If a pixel contains two or more particles, their energy and momenta are summed and stored as pixel values. The transformation of the jet representations is done on both the real and the test datasets. We label the ,,real'' jet images with the digit 1 and ,,test'' jet images with the digit 0. The classifier is then optimized by minimizing the \textit{binary crossentropy} loss between the real and the test datasets. The optimization is performed by ADAM algorithm \cite{adam}. It is important to note that the sizes of both datasets need to be the same.  
\subsubsection{Optimizing the neural network $f$}
After the classifier is optimized, a new jet dataset is generated by using the neural network $f$. Just as earlier, the generated jets are first transformed to jet images and then fed to the classifier. Since we have access to each of the decay probabilities for each jet, the right side of Eq.\ref{log_pq_jets} can be easily calculated for all the jet vectors $\vec{x}$ in the dataset. This way we can obtain the desired log value of the total probability for each jet $p(\vec{x})$:
\begin{equation}
\ln p(\vec{x}) = \ln{C_{NN}(\vec{x})} - \ln({1 - {C_{NN}(\vec{x})}})  + \sum_i^{2^N-1} \ln q(m_1^i, m_2^i | M).
\label{p}
\end{equation}
Finally, we update the parameters of the neural network $f$ by minimizing the expression given by:
\begin{equation}
L = \frac{1}{n} \sum_i^n \left[ \sum_{j}^{2^N-1} f(a_i^j,b_i^j,M_j) - \ln p_i(\vec{x})\right]^2,   
\label{loss}
\end{equation}
where $i$ denotes the jet index and $j$ denotes the decay index in a particular jet. After this step, the weights of the neural network are updated in such a way that the network output values $f(a,b,M)$ are on average closer to the real log value of $p(m_1,m_2 \mid M)$. The updated network $f$ is used to generate the test dataset in the next iteration. 

\subsection{Evaluation of the 2NN algorithm}
Upon completion of each iteration of the algorithm, the underlying probability densities can be obtained from the output values of the neural network $f$ according to Eq.\ref{q}. In the Results section the 2NN algorithm is evaluated in terms of the Kullback-Leibler divergence (KL) in the following way \cite{KLD}:
\begin{equation}
KL(M)  = \sum_{j} \sum_{k} p_{\textrm{real}} (m_1^j, m_2^k \mid M)\left[
\ln p_{\textrm{real}} (m_1^j, m_2^k \mid M) - f(a^j, b^k, M) + \ln{Z(M)}\right]
 \label{kl}
\end{equation}
where the sum is performed over the whole probability space. The KL-divergence is a non-negative measure of the difference between two probability densities defined on same probability space. If the probability densities are identical, the KL divergence is zero. 

\subsection{Hardware and software}
Code for calculations in this reasearch are is written in Python programming langauge using \textit{Tensorflow 2} and \textit{Numpy} modules. GPU unit NVIDIA Quadro p6000 obtained from the NVIDIA Grant for academic resarch is used to increase the speed of the performed calculations. 

\section{Results}
\begin{figure}[h!t!]
\centering
\includegraphics[width=15cm]{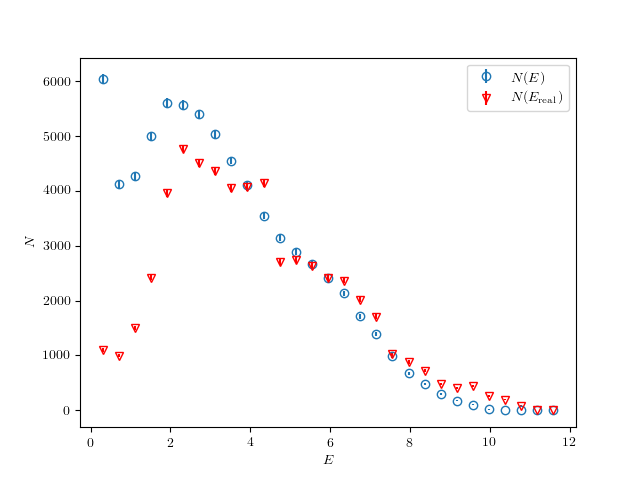}
\caption{The energy spectrum of the particles in the final state in jets generated by the calculated probabilities, compared to the energy spectrum of particles taken from jets belonging to the ,,real'' dataset.}
\label{hE}
\end{figure}

In this section we present our findings after applying the $2NN$ algorithm on 500 000 jets created using the particle generator described in \ref{ParticleGenerator}. In each iteration, the classifier is optimized using 50 000 randomly picked jets from the ,,real'' dataset and 50 000 jets generated using the neural network $f$. To optimize the neural network $f$, we use 50 000 jets as well. The algorithm performed 800 iterations. After the final iteration of the $2NN$ algorithm we obtain the calculated probability densities, which can be then used to generate samples of jets. First, we show the energy spectrum of the particles in the final state in jets generated by the calculated probabilities in \ref{hE}. This spectrum is directly compared to the energy spectrum of particles taken from jets belonging to the ,,real'' dataset and shown on Figure \ref{hE}.

The plotted spectra are obtained using 10 000 jets from each dataset. The error bars in the histogram are smaller than the marker size and are hence not visible. A resemblance between the two spectra is notable, especially at higher energies. This points to the fact that the calculated probabilities are approximately correct, so we can use them to generate samples of jets that resemble ,,real'' jets. To further examine the calculated probability densities we need to reconstruct the hidden resonances which are not found in the final state. For this purpose, the calculated probability densities for mother particle masses of $M = 25.0$, $M = 18.1$, $M = 14.2$ and $M = 1.9$ are analyzed and compared to the real probability densities in the following subsections. These masses are chosen since they match the masses of the hidden resonances, as was introduced in table \ref{TableParticles}. 

\subsection{Mother particle with mass $M$ = 25.0}

\begin{figure}[h!t!]
    \centering
    \begin{subfigure}{0.3\textwidth}
 
    \includegraphics[width=\linewidth]{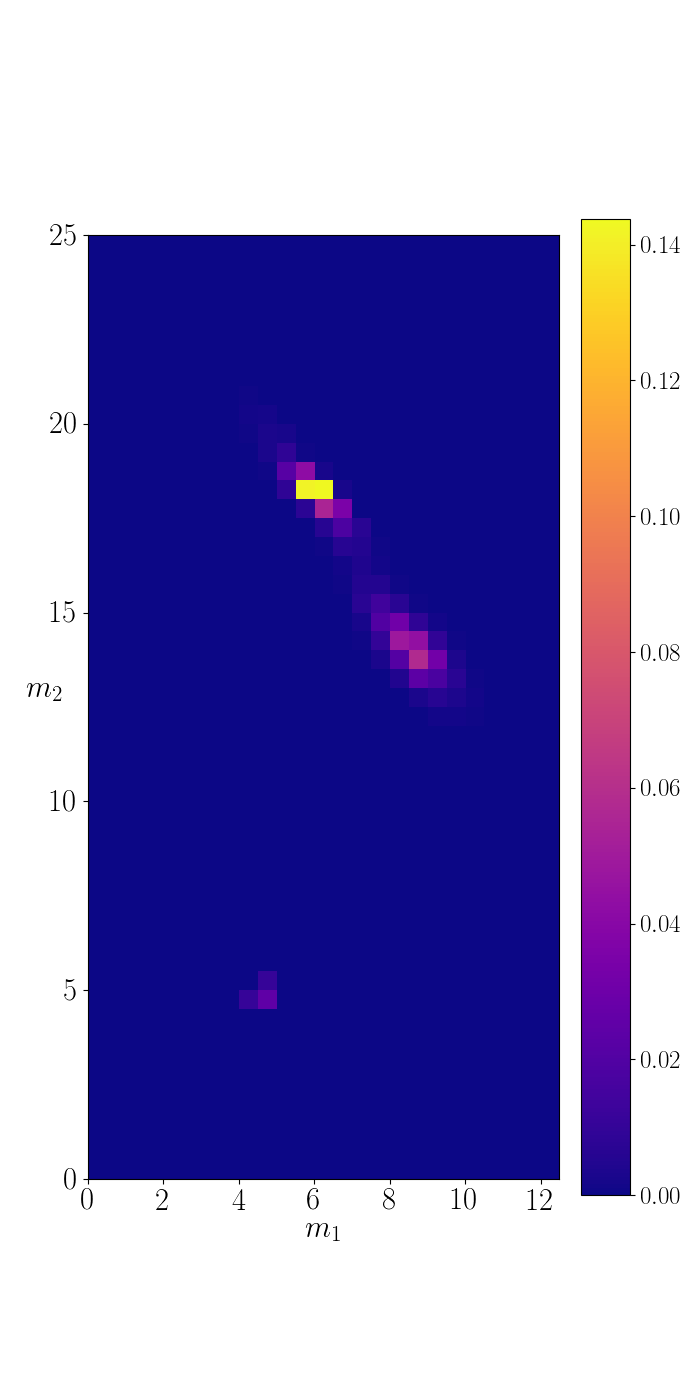}
       \caption{}
    \end{subfigure}
    \begin{subfigure}{0.3\textwidth}
    \includegraphics[width=\linewidth]{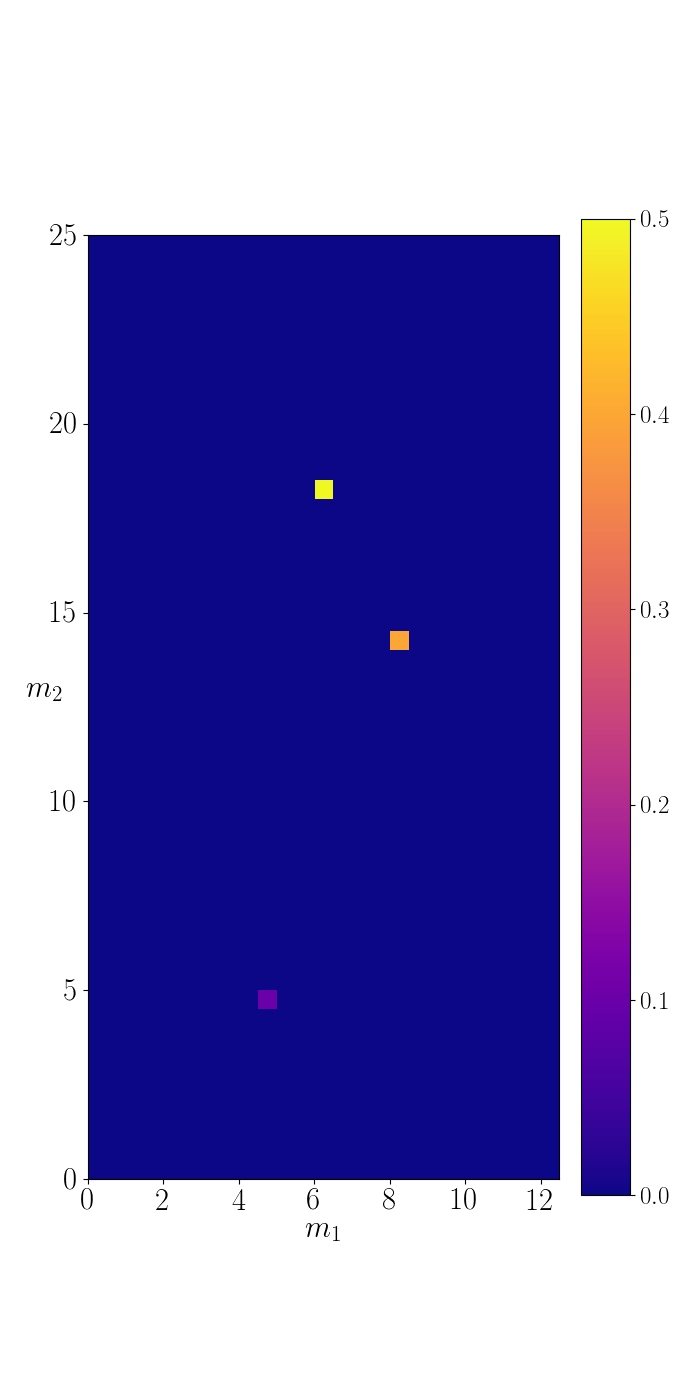}
       \caption{}
    \end{subfigure}
    \begin{subfigure}{0.3\textwidth}
    \includegraphics[width=\linewidth]{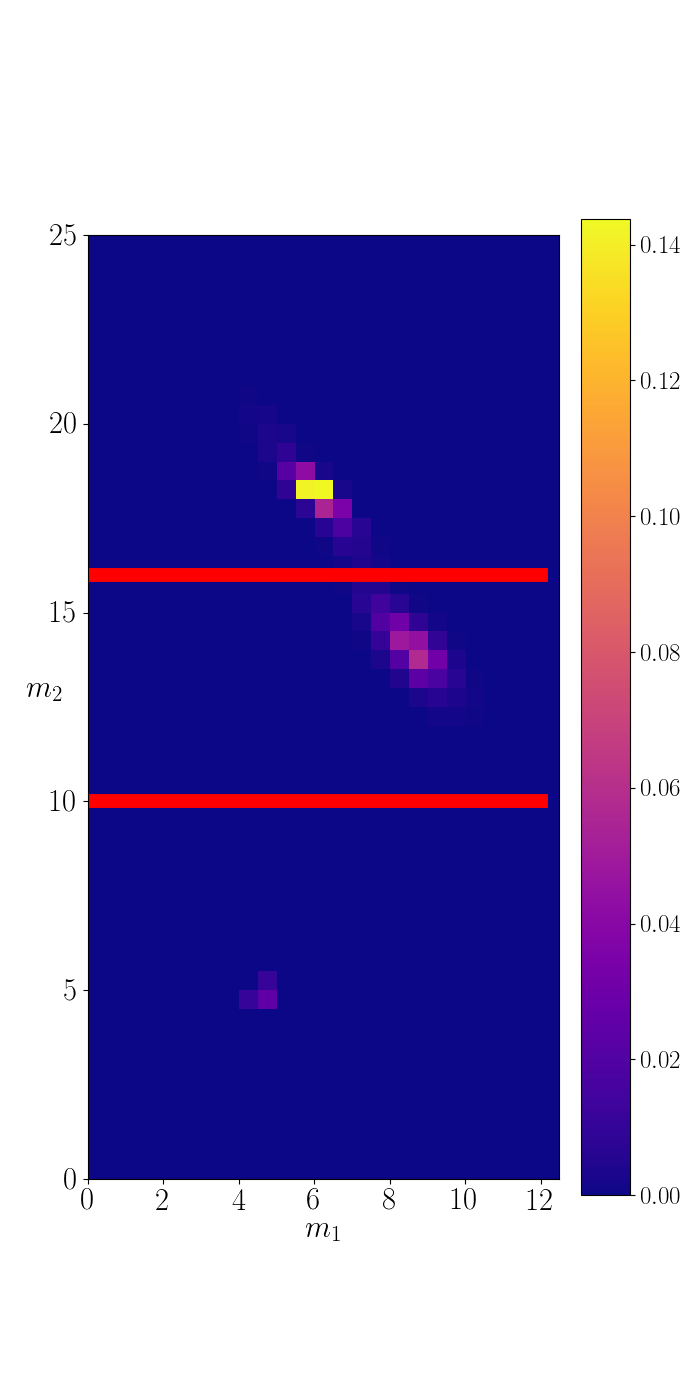}
       \caption{}
    \end{subfigure}
    \caption{The calculated probability density for a decaying particle of mass $M = 25.0$. (\textbf{a}) The left panel shows the density evaluated on the entire discretized probability space. (\textbf{b}) The probability density of ,,real'' data. (\textbf{c}) A division of the probability space into three subspaces, in order to isolate particular decays.}
    \label{probs25}
\end{figure}

The calculated 2$d$-probability density $p(m_1,m_2 \mid M)$ is shown in Figure \ref{probs25}, compared to the real probability density. A simple eye reveals that 3 possible decays of particle of mass $M = 25.0$ are recognized by the algorithm. After dividing the probability space as in panel (c) in Figure \ref{probs25} with lines $m_2 > 16.0$ and $m_2 < 10.0$, we calculate the mean and the variance of the data on each subspace. As a result, we obtain $(m_1, m_2) = (18.1 \pm 0.5, 6.1 \pm 0.5)$ for $m_2 > 16.0$, $(m_1, m_2) = (14.0 \pm 0.7, 8.4 \pm 0.7)$ for $16.0 \leq  m_2 > 10.0$ and  $(m1, m2) = (4.8 \pm 0.2, 4.6 \pm 0.2)$ for $m_2 \leq 10.0$. These mean values closely agree with the masses of the resonances expected as the products of decays of the particle with mass $M = 25.0$. The calculated small variances indicate that the algorithm is very precise. The total decay probabilities for each of the subspaces are equal to $p_1 = 0.48$, $p_2 = 0.47$, $p_3 = 0.05$, which approximately agree with the probabilities of decay channels of the particle with mass $M = 25.0$, as defined in table \ref{TableParticles}.

These results show that we can safely assume that the $2NN$ algorithm successfully recognizes all the decay modes of the particle that initiates a jet. To quantify the difference between the calculated probabilty density and the real probability density, we use the KL-divergence. 

\begin{figure}[h!t!]
\centering
\includegraphics[width=13cm]{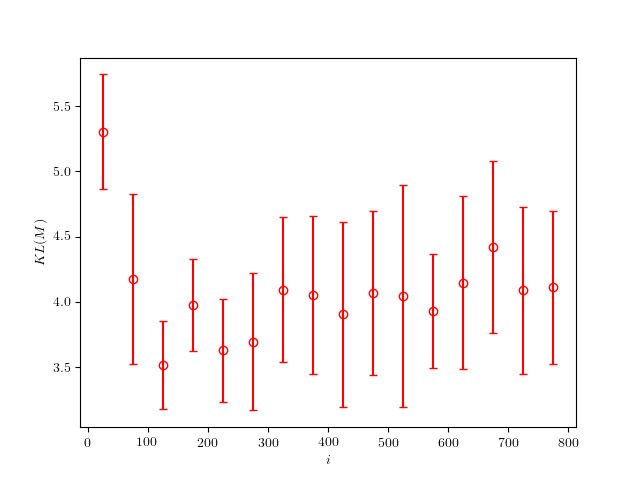}
\caption{The KL-divergence between the calculated and the real probability densities, evaluated in the case of particle of mass $M = 25.0$. The presented results are averaged over 50-iteration intervals. The error bars represent the standard deviation calculated on the same intervals.}
\label{kl25}
\end{figure}

Figure \ref{kl25} shows the dependence of the KL-divergence on the iteration of the $2NN$ algorithm. First, we observe an initial steep decrease in the value of the divergence. Large variations in divergence value are observed later. This is an indicator that the approximate probability density is found relatively quickly - after a few hundred iterations. As the algorithm decreases the width of the peaks found in the probability distribution, the KL-divergence becomes very sensitive to small variations in the location of these peaks and can therefore vary by a large relative amount.  

\subsection{Mother particle with mass $M$ = 18.1}

\begin{figure}[h!t!]
    \centering
    \begin{subfigure}{0.45\textwidth}
    \includegraphics[width=\linewidth]{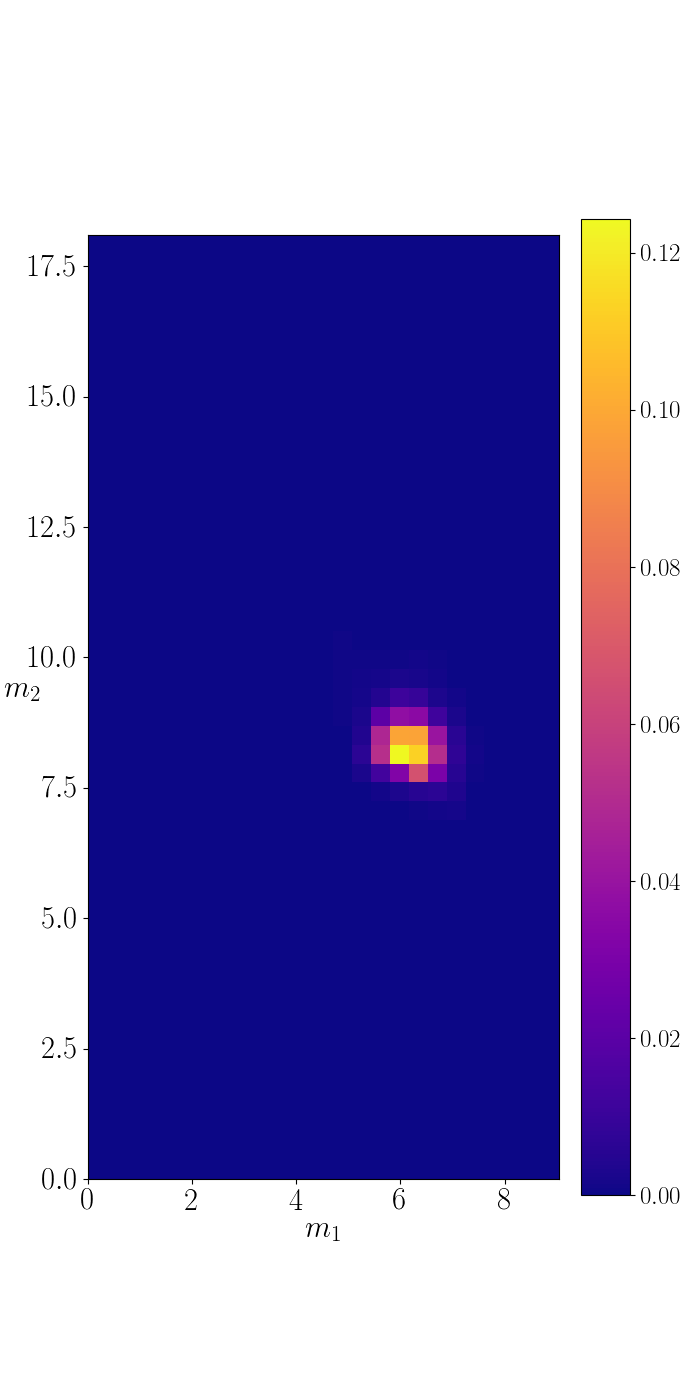}
    \end{subfigure}
    \begin{subfigure}{0.45\textwidth}
    \includegraphics[width=\linewidth]{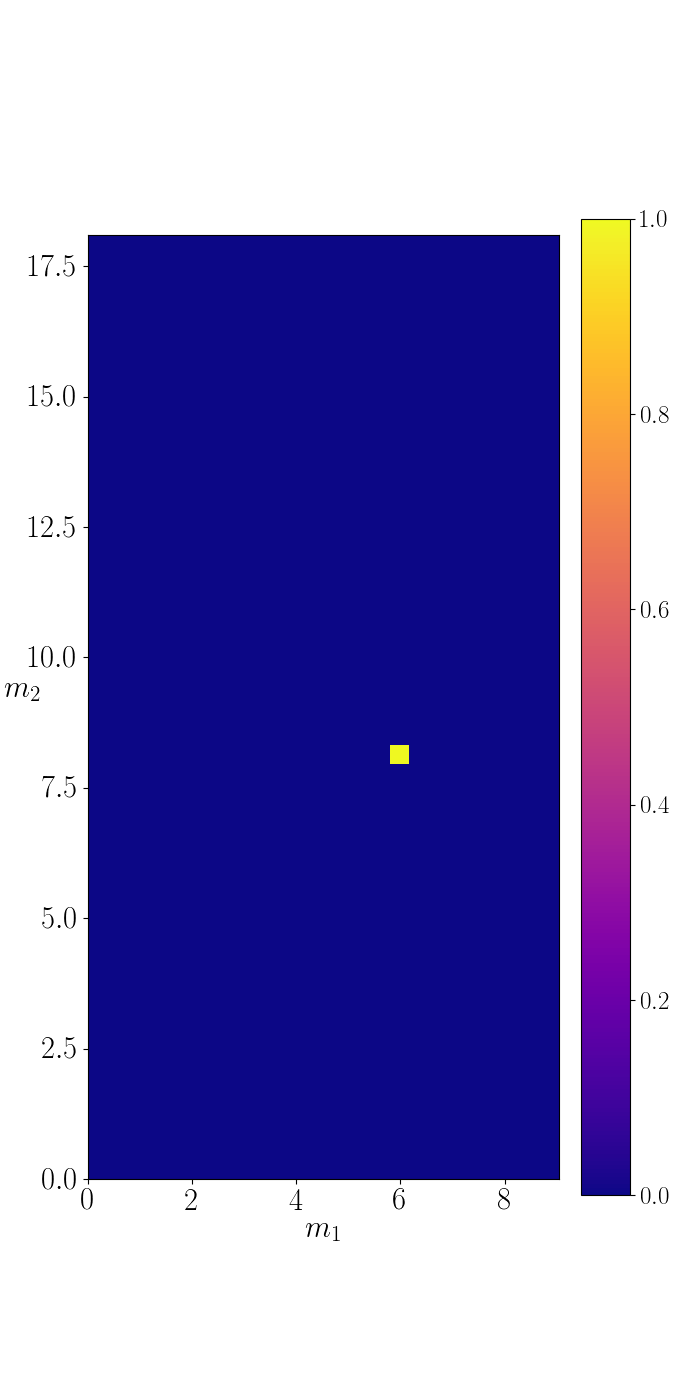}
    \end{subfigure}
    \caption{The calculated probability density for a decaying particle of mass $M = 18.1$. (\textbf{a}) The calculated density evaluated on the entire discretized probability space. (\textbf{b}) The probability density of ,,real'' data. }
    \label{probs18}
\end{figure}

A similar analysis is performed for the particle with mass $M = 18.1$. The calculated probability density is shown in Figure \ref{probs18} compared to the expected probability density. In this case, only one decay is allowed, so a division into probability subspaces is not necessary, as was for the case when $M$=25.0. The calculated mean and the variance of the shown probability density are $(m_1, m_2) = (5.9 \pm 0.4, 8.2 \pm 0.6)$. In this case, just as in the former, the calculated values closely agree with the only possible decay, in which the mother particle decays into two particles of masses 6.1 and 8.4.  Also, just as in the previous subsection, the obtained result is very precise. Therefore, the algorithm can successfully find hidden resonances, as well as recognize the decay channels, without ever seeing them in the final state in the ,,real'' dataset. 

\begin{figure}[h!t!]
\centering
\includegraphics[width=13cm]{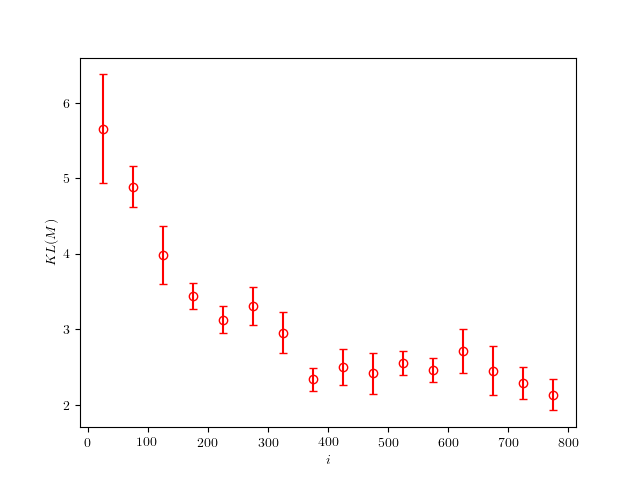}
\caption{The KL-divergence between the calculated and the real probability densities, evaluated in the case of particle of mass $M = 18.1$. The presented results are averaged over 50-iteration intervals. The error bars represent the standard deviation calculated on the same intervals.}
\label{kl18}
\end{figure}

The calculated KL-divergence in the case of particle with mass $M = 18.1$ decreases over time in a very smooth manner, as can be seen in Figure \ref{kl18}. We believe this could be due to the simpler expected probability density, which algorithm manages to find very quickly.  

\subsection{Mother particle with mass $M$ = 14.2}
\begin{figure}[h!t!]
    \centering
    \begin{subfigure}{0.3\textwidth}
 
    \includegraphics[width=\linewidth]{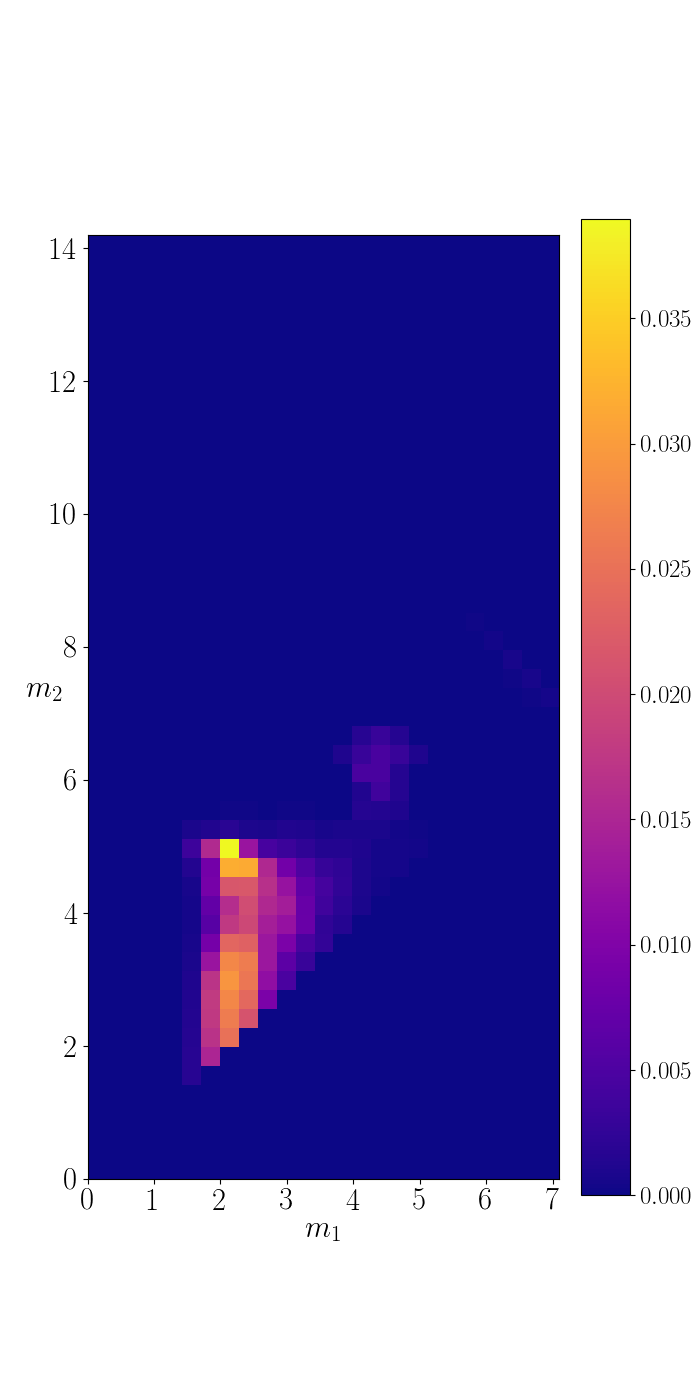}
       \caption{}
    \end{subfigure}
    \begin{subfigure}{0.3\textwidth}
    \includegraphics[width=\linewidth]{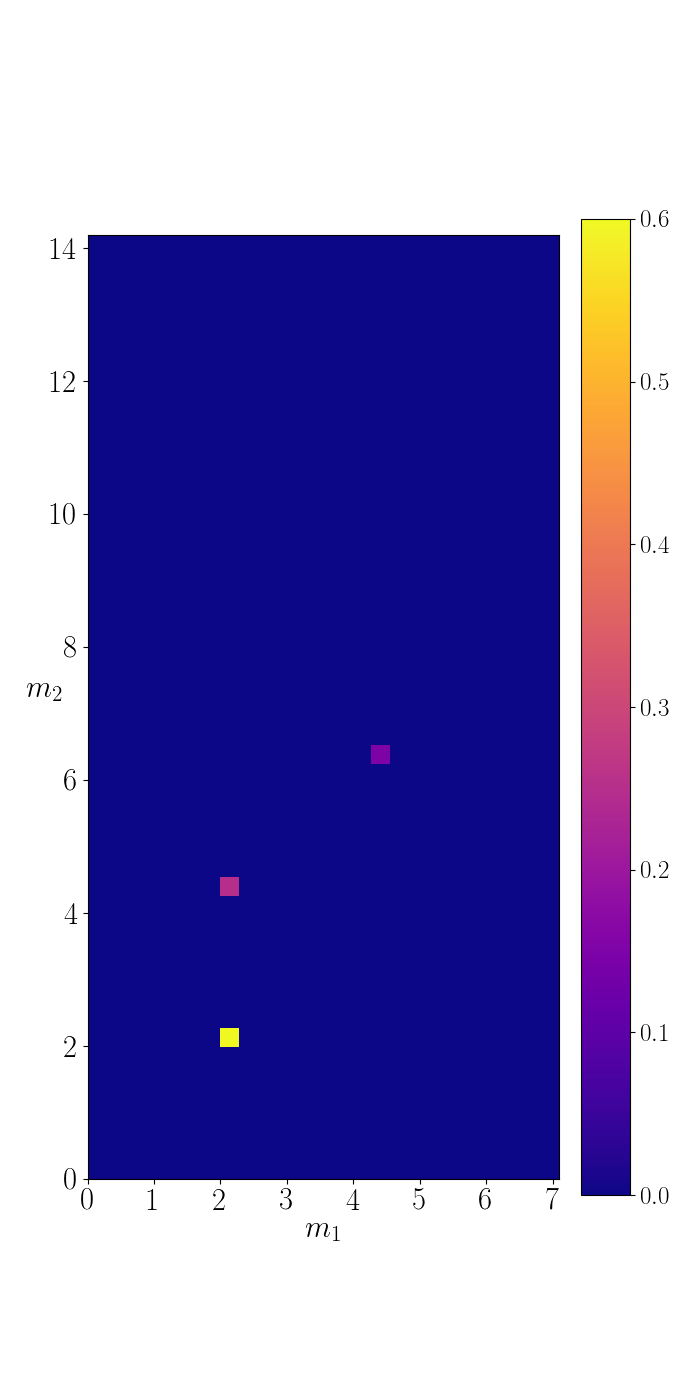}
       \caption{}
    \end{subfigure}
    \begin{subfigure}{0.3\textwidth}
    \includegraphics[width=\linewidth]{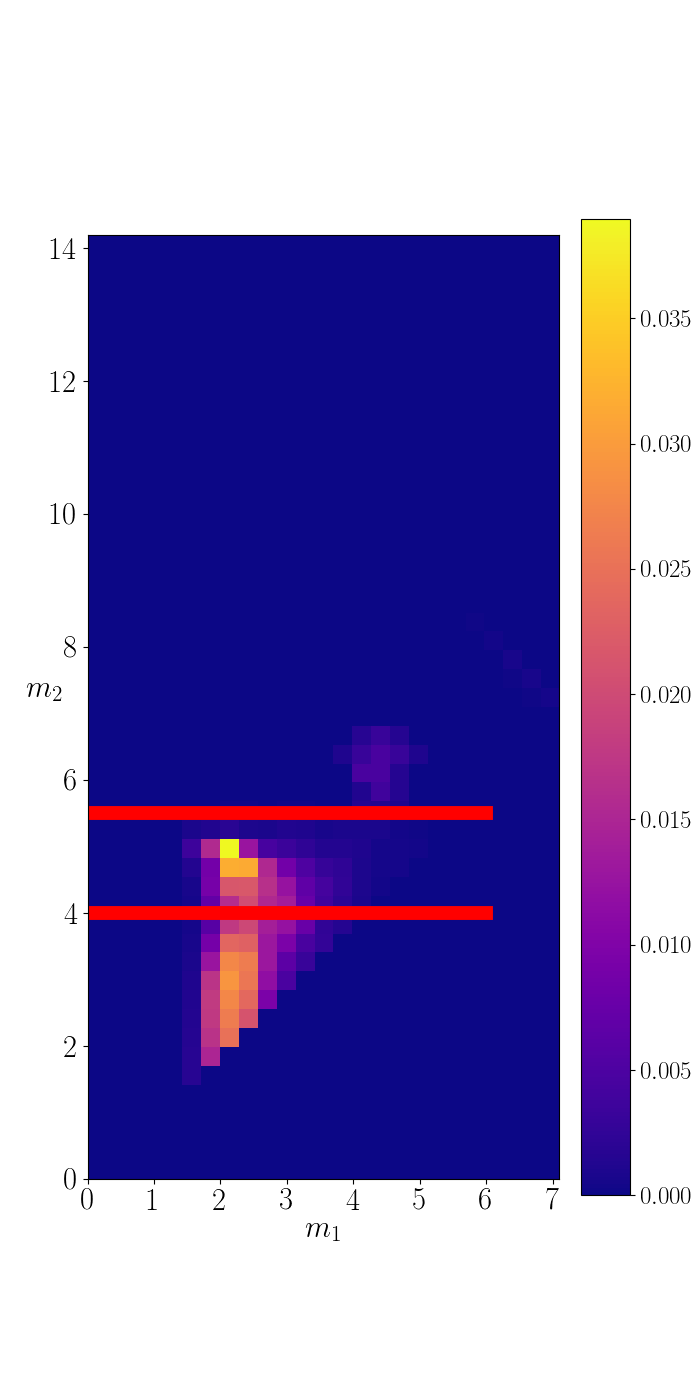}
       \caption{}
       \label{probs2c}
    \end{subfigure}
    \caption{The calculated probability density for a decaying particle of mass $M = 14.2$. (\textbf{a}) The left panel shows the density evaluated on the entire discretized probability space. (\textbf{b}) The probability density of ,,real'' data. (\textbf{c}) A division of the probability space into three subspaces, in order to isolate particular decays.}
    \label{probs14}
\end{figure}

\begin{figure}[h!t!]
\centering
\includegraphics[width=13cm]{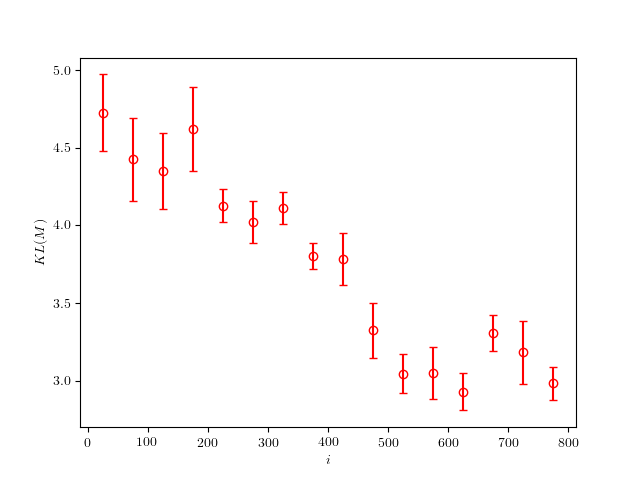}
\caption{KL-divergence between calculated and real probability density evaluated for the $M = 14.2$. Results are averaged over the intervals of 50 iteration. Error bars represent standard deviation on the same interval}
\label{kl14}
\end{figure}

Figure \ref{probs14} shows the 2$d$-probability density for the decaying particle of mass $M = 14.2$. In this case, we can identify 3 possible decay channels, which are not as clearly separated as the channels in the previous subsections. Similar to the case of decaying particle of mass $M = 25.0$, we divided the probability space into 3 subpaces, each of which covered one of the possible decays. In this case, the three subspaces cover areas where $m_2 \leq 4.0$, $4.0 < m_2 \leq 5.5 $ and $m_2 > 5.5$. The mean values of the probability density on each of the subspaces are $(m_1,m2) = (2.4 \pm 0.5, 2.9 \pm 0.7)$, $(m_1,m_2)= (2.7 \pm 0.7, 4.3 \pm 0.3)$ and $(m_1,m_2) = (4.4 \pm 0.4, 6.2 \pm 0.3)$, respectively. The allowed decays of a mother particle with mass $M$ = 14.2 in the ,,real'' data are into channels with masses $(1.9,1.9)$, $(1.9, 4.4)$ and $(4.4, 6.2)$, which agree with the calculated results. However, in this case the calculations show higher variance, especially for decays where one of the products is a particle with mass 1.9. The total probabilities of decay in each of the subspaces are 0.89, 0.05 and 0.06, respectively. The relative probabilities of decay channels into particles with masses (4.4, 6.1) and (1.9, 4.4) are approximately the same as expected. However, the algorithm predicts more decays in the channel (1.9,1.9) than expected. The KL-divergence shows a steady decrease with occasional spikes, as shown on Figure \ref{kl14}.

\subsection{Mother particle with mass $M$ = 1.9}
\begin{figure}[h!t!]
    \centering
    \begin{subfigure}{0.3\textwidth}
    \includegraphics[width=\linewidth]{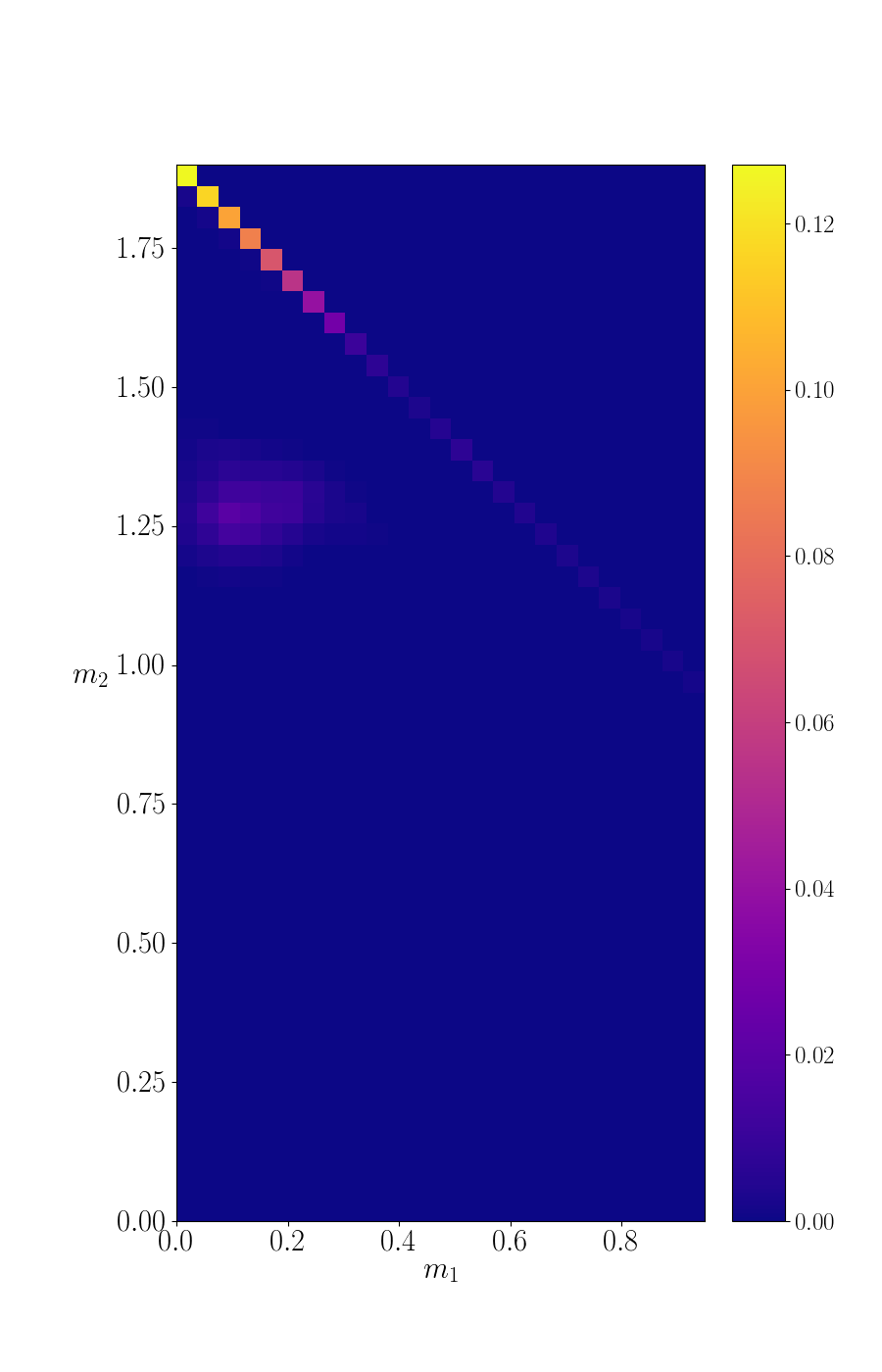}
       \caption{}
    \end{subfigure}
    \begin{subfigure}{0.3\textwidth}
    \includegraphics[width=\linewidth]{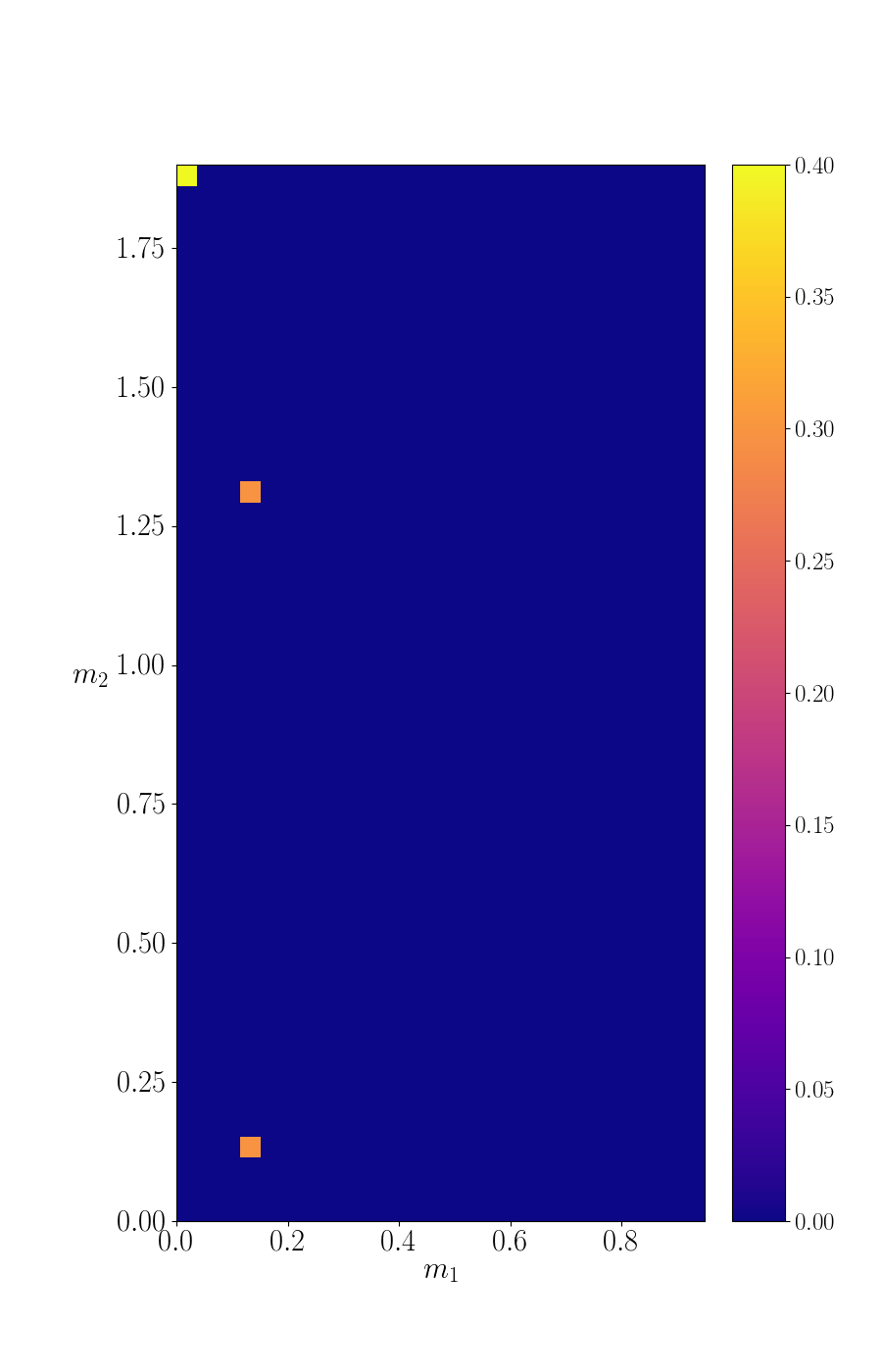}
       \caption{}
    \end{subfigure}
    \begin{subfigure}{0.3\textwidth}
    \includegraphics[width=\linewidth]{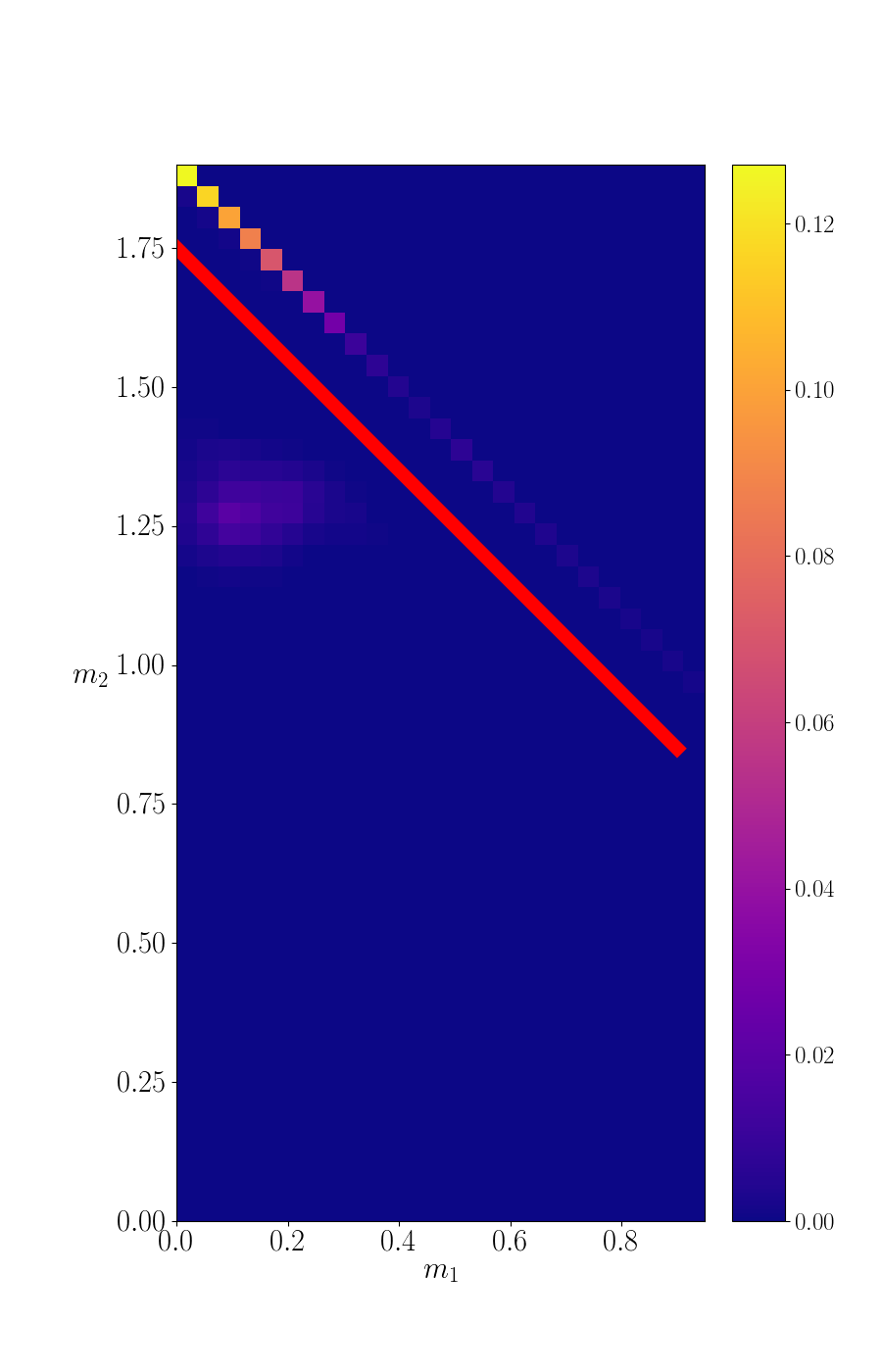}
       \caption{}
    \end{subfigure}
    \caption{The calculated probability density for a decaying particle of mass $M = 1.9$. (\textbf{a}) The left panel shows the density evaluated on the entire discretized probability space. (\textbf{b}) The probability density of ,,real'' data. (\textbf{c}) A division of the probability space into three subspaces, in order to isolate particular decays.}
    \label{probs2}
\end{figure}

The last probability density we analyze is the probability density for the mother particle with mass $M$ = 1.9. Figure \ref{probs2} shows the calculated probability density. It can be seen that one of the decay modes present in the ,,real'' data, namely when the particle decays in the $(0.1, 0.1)$ channel, is not recognized by the algorithm, but the decay mode when the particle decays in the $(0.1, 1.3)$ channel is visible. If we isolate given decay as shown in the right panel of Figure \ref{probs2}, we get a mean value of $(m_1, m_2) = (0.14 \pm 0.09, 1.27 \pm 0.09)$, which agrees with the expected decay. We also observe significant decay probabilities along the line $m_1 + m_2 = 1.9$. The decays that correspond to the points on this line in effect create particles with zero momentum in the rest frame of the mother particle. In the lab frame this corresponds to the daughter particles flying off in the same direction as the mother particle. Since they reach the detector in the same time, they are registered as one particle of total mass $M = 1.9$. Thus, we can conclude that the probabilities on this line have to add up to the total probability of the mother particle not decaying. The calculated probabilities in the case of no decay and in the case when decaying into particles with masses $(0.1,1.3)$ are 0.71 and 0.29, respectively. We note that relative probabilities are not correct, but 2 of the 3 decay modes are still recognized by the algorithm. The KL-divergence in this case can't produce reasonable results, simply because of multiple points in the $(m_1,m_2)$ phase space which produce the same decay and is therefore omitted from the analysis.

\subsection{The accuracy of the classifier}
\begin{figure}[h!t!]
\centering
\includegraphics[width=13cm]{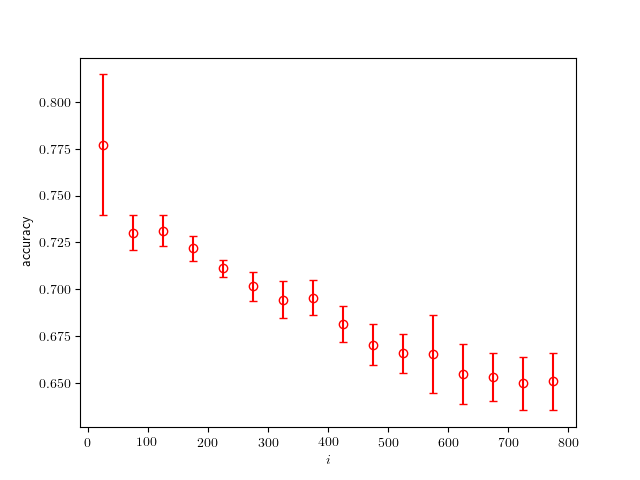}
\caption{The calculated accuracy of the classifier in dependence on the iteration number.}
\label{acc}
\end{figure}

The accuracy of the classifier is defined as the fraction of correctly ,,guessed'' samples on a given dataset. The criterion used for guessing is checking whether the output of the classifier, $C_{NN}$, is greater than 0.5. The accuracy can indirectly indicate how distinguishable are some two datasets. In our algorithm, after starting from a test probability density, we approach the real probabilility density with increasing iteration number, so we can expect that the two jet datasets, the ,,real'' and the ,,test'' dataset, are less and less distinguishable over time. In Figure \ref{acc} we show the accuracy of the classifier in dependence on the iteration number. 

After an initially high value, the accuracy decreases with growing iteration number, which demonstrates that the test dataset becomes more and more similar to the real dataset. Ideally, the datasets are no longer distinguishable by a given classifier if the evaluated accuracy reaches 0.5. Therefore, we can use the evaluated accuracy of the classifier as a criterion for stopping the algorithm. Other measures can also be used as the stopping criterion such as the loss value of the classifer or the area under reciever operating characteristic (ROC) curve of the classifier. In this work, the algorithm is stopped after the accuracy reaches a value of 0.65, because we didn't see any significant decreasy in the accuracy once it reached this value. An accuracy value of 0.65 clearly shows that the classifier is capable of further discriminating between the two datasets. This is explained by the fact that the neural network $f$ and its hyperparameters are not fully optimized. For the algorithm to perform better, we need to optimize the neural network $f$ and possibly improve the architecture for the selected task.   

\newpage
\section{Discussion}
In this work we propose a method for calculating underlying probability distributions in particle decays, using only the data that can be collected in a real-world physical system. First, we developed an artificial physical system based on the QCD fragmentation process. Next, we present the core part of the method: the $2NN$ algorithm, which we described in detail. The algorithm performs very well when tested on the developed physical system. It accurately predicts most of the hidden resonant particles, as well as their decay channels, which can occur in the evolution of jets. The energy spectra of the particles in the final state can also be accurately reproduced. 

Although tested only on the developed artificial physical system, we believe that the method is general enough to be applicable to real-world physical systems, such as collisions of high-energy particles, with a few possible modifications. For example, we hope that this method can in the future prove helpful in measuring the fragmentation functions of quarks and gluons. Also, one could employ such a method in the search for supersymmetric particles of unknown masses, or in measuring the branching ratios of known decays.  

The $2NN$ algorithm does not specify the exact architecture of the neural networks, nor the representation of the data used. Furthermore, the classifier does not need to be a neural network - it can be any machine learning technique which maximizes likelihood. Although the algorithm has a Generative Adversarial Netowrk (GAN)-like structure, it converges readily and does not show usual issues associated with  GANs, such as mode collapse or vanishing gradients. The downside of the presented algorithm are high computational requirements. Continuous probability distributions, which we expect to occur in nature, are approximated by discrete probability distributions. In quest for higher precision and a better description of reality, one always aims to increase the resolution of discrete steps, but this carries a high computational cost. Also, the used neural networks are not fully optimized, which slows down the convergence of the algorithm. In conclusion, in order to cut down computational costs, a more thorough analysis of convergence is needed to achieve better performance.  

In future work we hope to make the method even more general and thus even more applicable to real-world physical systems. In particular, we want to introduce angle dependent probability distributions, which can be retrieved from some detector data. We would also like to investigate the possibility of including other decay modes, such as $1 \rightarrow 3$ type decays. 

\newpage
\appendixtitles{yes} 

\appendix
\section{Description of the neural networks}
The classifier used to recover the underlying probability distributions is a feed forward convolutional neural network (CNN), already used in \cite{NNNP3}. In that work, the classifier was used for a different purpose, but its architecture proved to be suitable for the task at hand.

The neural network $f$ is a \textit{feed-forward} neural network consisting of 5 independent completely connected layers. Each of the layers consist of 100 neurons apart from the last one, which is a single neuron. The activation function used in all the layers, apart from the last one, is a ReLu activation function. The last layer has no activation function. The network was optimized using the ADAM algorithm.

\vspace{6pt} 

\authorcontributions{The authors contributed equally to all segments of this paper.}

\funding{This research was funded by the Croatian science foundation grant IP-2018-01-4108 ``Demystifying Two Particle Correlations in pp collisions with the upgraded Time Projection Chamber''.}

\acknowledgments{We gratefully acknowledge the support of NVIDIA Corporation with the donation of the Quadro P6000 graphics processing unit used for this research.}

\conflictsofinterest{The authors declare no conflict of interest.} 

\abbreviations{The following abbreviations are used in this manuscript:\\
\noindent 
\begin{tabular}{@{}ll}
QCD & Quantum Chromodynamics\\
LHC & Large Hadron Collider\\
CERN & Conseil Européen pour la Recherche Nucléaire\\
2NN & 2 Neural Networks\\
ROC & Reciever Operating Characteristic \\
GAN & Generative Adversarial Network \\
CNN & Convolutional Neural Network
\end{tabular}}

\reftitle{References}
\newpage
\raggedright


\end{document}